# Eye Pupil Location Using Webcam

Michal Ciesla*, Przemyslaw Koziol

*Department of Physics, Astronomy and Applied Computer Science, Jagiellonian University, Reymonta 4, 30-059 Kraków, Poland.*

## Abstract

Three different algorithms used for eye pupil location were described and tested. Algorithm efficiency comparison was based on human faces images taken from the BioID database. Moreover all the eye localisation methods were implemented in a dedicated application supporting eye movement based computer control. In this case human face images were acquired by a webcam and processed in a real-time.

keywords: eye-driven computer control, human computer interfaces, eye pupil detection.

*corresponding author: michal.ciesla@uj.edu.pl

# 1. Introduction

For most of us the sense of sight is the primary source of data about surrounding environment. Therefore it is natural to assume that information about where a gaze is focused could be helpful in determining how we communicate with the surroundings. In the area of Human-Computer Interaction (HCI) that knowledge is crucial for creating an intuitive and ergonomic user interface [1]. However the fundamental step in implementing such an interface is the exact location of a user eye pupil.

The history of eye tracking reaches back to the late 19[th] century. At the beginning, mechanical devices were used to detect light reflected by a plate implanted directly into the cornea. Development of photography and video recording allowed for much more reliable and less invasive methods of eye movement observation over long periods of time. Such studies became more popular specially in psychology and medical research as well as in diagnostics. However, it has been only recently that the computing power become high enough to allow for the development of a computer interface based on a real-time eye-tracking analysis.

Currently, the eye tracking techniques develop in two directions, electrooculography (EOG) and digital image analysis. The last one, which is the research area of this work, uses cameras operating in the visible light spectrum and software analyzing digital images. The increase in computing power also gave way to the number of techniques carrying out such analysis. The advantage of methods using visible light is their versatility. They are independent of such individual characteristics of an eye such as current flow in the cornea.
Unfortunately, the commercially available applications require specialized equipment (e.g. sensitive low-noise video camera allowing fast transfer of high resolution frames), which makes them quite expensive [2,3]. There is also alternative approach using open-source software based on eye pupil reflection in infrared light, but the hardware needed limits its versatility [4]. The only freeware solution using visible light is the EyeTrack [5]; however, it does not allow for the precise comparison of different algorithms.

The aim of this paper is to describe selected algorithms for eye pupil detection, compare their effectiveness using static digital images, and implement in an application for eye-controlled computer operation. The effectiveness assessment was based on the collection of facial images [6] with the actual pupils locations attached. The implementation was performed using an ordinary webcam, with a standard resolution of 640 x 480 or even 320x240 pixels.

In the next section, the general algorithm for finding eyes on a digital image is presented. The following section concentrates on three commonly used methods for determining eye pupil position. It also describes all the important implementation details, e.g. threshold values providing the best possible results. Section 4 provides the comparison of the algorithms performed on static images as well as on images acquired in a real-time mode. Short summary presents the conclusions.

## 2. General Algorithms for Eye-Control Computer Operation

Eye-driven computer operation requires certain steps of processing of an image captured with a video recording device. The diagram below presents the general scheme of the process.

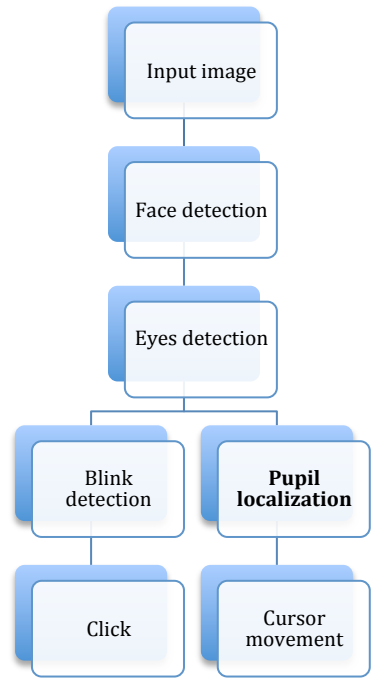

*Fig 1. General scheme of an image processing sequence used for eye driven computer control.*

Although all the steps are very important, here we would like to focus mainly on the pupil location methods. Other algorithms used in the prepared software package are discussed in other papers [7-10].

## 3. Eye Pupil Location Algorithms

In the section, the short description of the three most popular methods used for location of an eye pupil is presented. Although their authors have already described all of them, there are always aspects strongly dependent on the given set of input data that should be clarified before implementation stage. For the purposes of the paper, it is assumed that all human face images are converted to 8-bit grey scale.

### 3.1. Cumulative Distribution Function (CDF) Algorithm

The method is based on the observation that an eye iris and pupil is much dimmer than cornea. The algorithm was proposed by Asadifard and Shanbezadeh [11]. Its name

comes from the Cumulative Distribution Function (CDF) of eye luminance used in the algorithm:

$$CDF(r) = \sum_{w=0}^{r} p(w),$$

where $p(w)$ is probability of finding point having luminance equal to $w$.

In the first step the algorithm changes the intensity $I(x,y)$ of each pixel of the input image as follows:

$$I'(x,y) = \begin{cases} 255 & \text{if} \quad CDF(I(x,y)) < 0.05 \\ 0 & \text{otherwise} \end{cases}$$

Parameter 0.05 was chosen experimentally to provide the best possible results. Example of the transformation result is presented below.

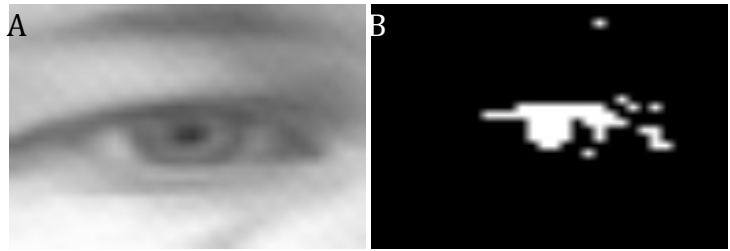

Fig 2. CDF filter: A) input image, B) filtered image.

The next step is the application of the minimum filter to remove singular white points and compact white region.

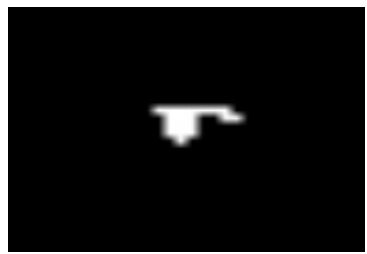

Fig 3. Application of the minimum filter with radius 2.

Then the algorithm chooses one white pixel, which is the darkest on the original input image. This pixel is called PMI (Pixel with Minimum Intensity).
As the probability that PMI belongs to an eye iris and not to a pupil is significant the further processing is needed. Therefore the algorithm returns to the original image and measures average intensity (AI) in 10x10-pixel square around PMI. Then the region is expanded to 15x15 pixels and minimum filter is applied. The eye centre is assumed to be a geometrical centre of points of intensity lower than AI calculated before.

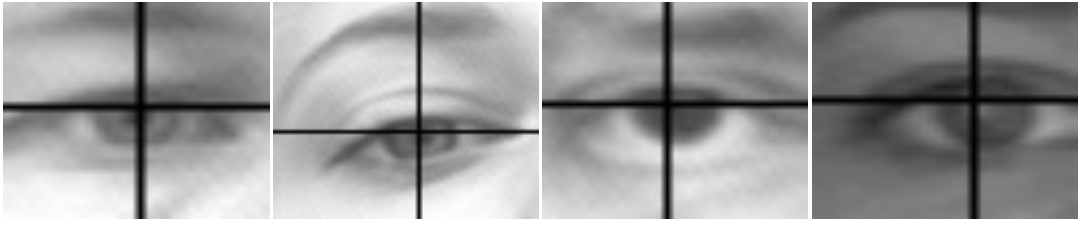

*Fig 4. Examples of eye pupil location using CDF algorithm.*

### 3.2. Projection Functions (PF) Algorithm

The idea of the method proposed by Zhou and Geng [12] is similar to the one used in CDF algorithm, but in this case pixel intensities are projected on vertical and horizontal axes. Those projections divide the whole picture to homogenous subsets – *Fig.5*. Division points $\{x_1, x_2, x_3, x_4\}$ and $\{y_1, y_2\}$ are connected with rapid change of the given projection function *PF* (horizontal or vertical):

$$\{x_1, x_2, x_3, x_4\} = \left\{x: \left|\frac{d\,PF_v(x)}{dx}\right| > T\right\}, \quad \{y_1, y_2\} = \left\{y: \left|\frac{d\,PF_h(y)}{dy}\right| > T\right\},$$

where *T* is an arbitrary chosen threshold value.

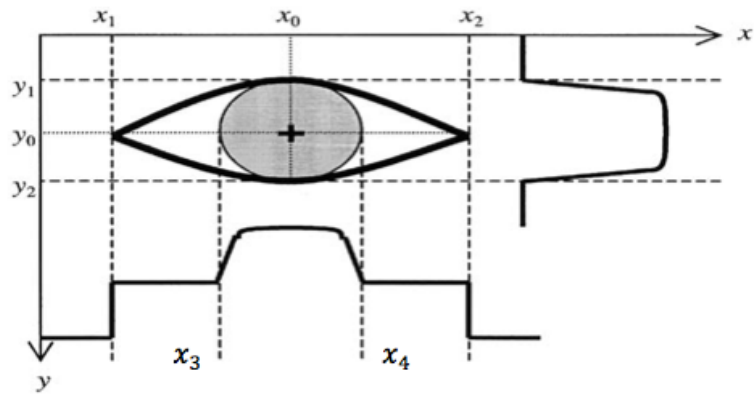

*Fig 5. . Projection Functions and their relation to pupil position.*

Pupil position $(x_0, y_0)$ is determined as following:

$$x_0 = \frac{x_3 + x_4}{2}, \quad y_0 = \frac{y_1 + y_2}{2}.$$

Values $x_1$ and $x_2$ are not taken into account as they do not provide any information about changes of pupil position in relation to eye corners.
The effectiveness of presented method depends on the specific definition of a projection function. The most popular options are the Integral Projection Function (IPF) and the Variance Projection Function (VPF):

$$IPF_h(y) = \frac{1}{x_b-x_a}\sum_{x=x_a}^{x_b} I(x,y) \qquad IPF_v(x) = \frac{1}{y_b-y_a}\sum_{y=y_a}^{y_b} I(x,y)$$

$$VPF_h(y) = \frac{1}{x_b-x_a}\sum_{x=x_a}^{x_b} \big(I(x,y) - IPF_h(y)\big)^2 \qquad VPF_v(x) = \frac{1}{y_b-y_a}\sum_{y=y_a}^{y_b} \big(I(x,y) - IPF_v(x)\big)^2$$

However, the best results are obtained using the General Projection Function (GPF):

$$GPF_h(y) = (1-\alpha)IPF_h(y) + \alpha VPF_h(y), \quad GPF_v(x) = (1-\alpha)IPF_v(x) + \alpha VPF_v(x),$$

with parameter $0 \leq \alpha \leq 1$. Zhou and Geng proved experimentally that the optimal value of $\alpha$ is 0.6, whereas in our tests the best results were obtained for $\alpha = 0$.

The following figures illustrate the process of determining projection functions and its efficiency in finding the centre of an eye pupil.

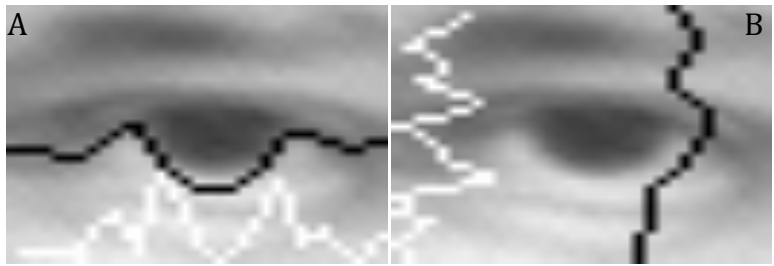

Fig 6. The plot of vertical (A) and horizontal (B) General Projection Function (black) and its derivative (white) over a grey scale picture acquired with webcam.

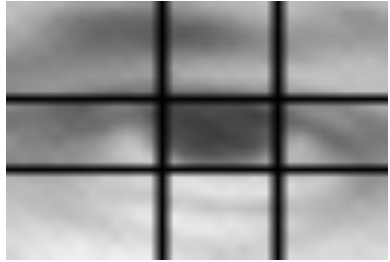

Fig 7. Edges of iris found on a picture presented in Fig. 6.

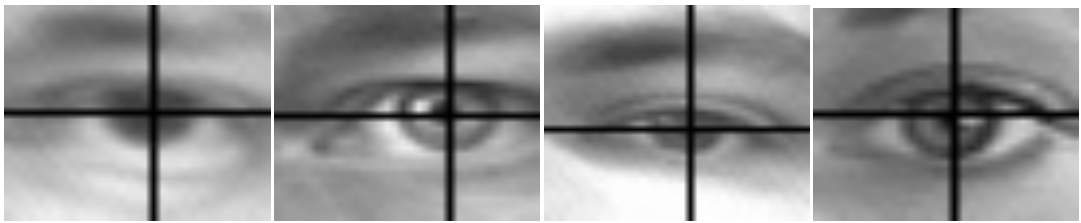

Fig 8. Examples of a pupil location using projection functions.

## 3.3. Edges Analysis (EA)

The method originates from the work of S. Asteriadis, et. al. [13], in which the edge pixel information was used for eye location in a picture of a human face. The input frame is processed by the most popular edges detection algorithm for digital images developed by Canny [14], however before that the Gaussian blur filter is applied to eliminate the undesired noise. The Canny method is based on two threshold values, upper and lower. The upper threshold value defines the minimum gradient needed to classify pixel as an edge component. Such a pixel is also called strong edge pixel. In the edge, there are also pixels of a gradient between the upper and lower threshold values, having at least one strong edge pixel as a neighbour. The lower threshold protects against splitting edges in low contrast regions.
In our work the lower and upper threshold values were set to 1.5 and 2.0 times the mean luminosity, respectively. The output of the Canny method is a binary picture with edges marked white (see fig. 9).

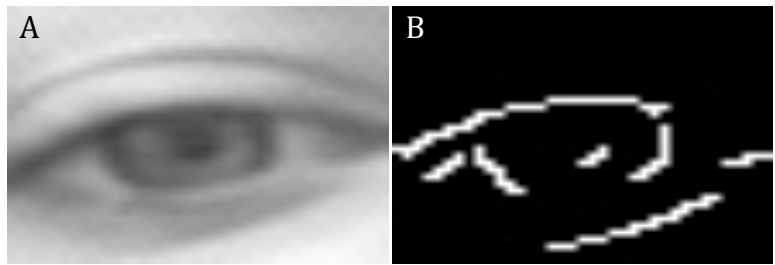

*Fig 9. Input image (A) and the processing result of Canny algorithm (B); edges are coloured white.*

The next step of the pupil detection process is to find vertical and horizontal lines sharing the next to highest number of points with the edges. The intersection of the lines indicates the pupil centre [13]. Unfortunately, the efficiency of this method was not satisfactory in our case. Therefore we modified it having observed that the vertical lines of the highest number of pixels shared with the edges cross the left and right iris-cornea boundary. Similar horizontal lines pass across the upper and lower border between iris and eyelid (see Fig.10). Additionally the modified method requires the lines to be at least 7 pixels apart (for an eye region of the approximate size of 30x30 pixels) to avoid artefacts occasionally appearing on webcam frames.

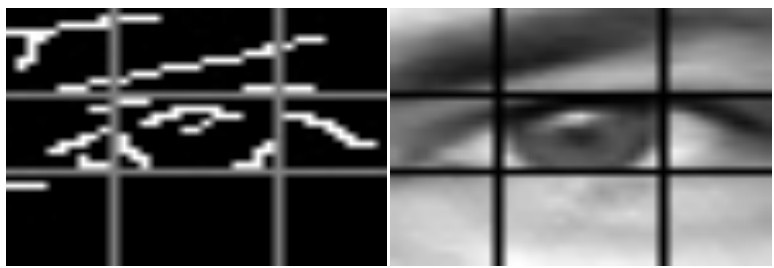

*Fig 10. Example of horizontal and vertical lines calculated by the modified edge analysis algorithm.*

Having boundary lines, the centre of an eye pupil is calculated in the same way as with the PF algorithm described in the previous section.

## 4. Results

### 4.1 Comparison using static images

Algorithms described in Section 3 were tested on the BioID databsase [6]. It contains 1521 grey level images of 384x286-pixel resolution. In all the images faces of 23 different test persons are presented en face, one face per image. Images vary in terms of background, illumination and scale. All of them contain information of the actual eye positions stored in additional file.

### 4.1.1 Detection error

Detection error describes the accuracy of eye pupil location algorithm. It is defined as [11]:

$$d = \frac{max\{\|L - L'\|, \|R - R'\|\}}{L - R},$$

where L, R are the actual positions of left and right pupil, respectively, while L' and R' are positions calculated by the tested algorithm. The above equation can only be used when both eyes regions are properly determined. For this purpose the OpenCV [15] implementation of the Viola-Jones method [7] was used. It turned out to be successful for 941 out of 1521 images. Therefore the efficiency of eye pupil location algorithm for a given detection error $d_{max}$ is defined as a number of images for which the method provides $d < d_{max}$ divided by 941. Figure 11 and Table 1 present the obtained results.

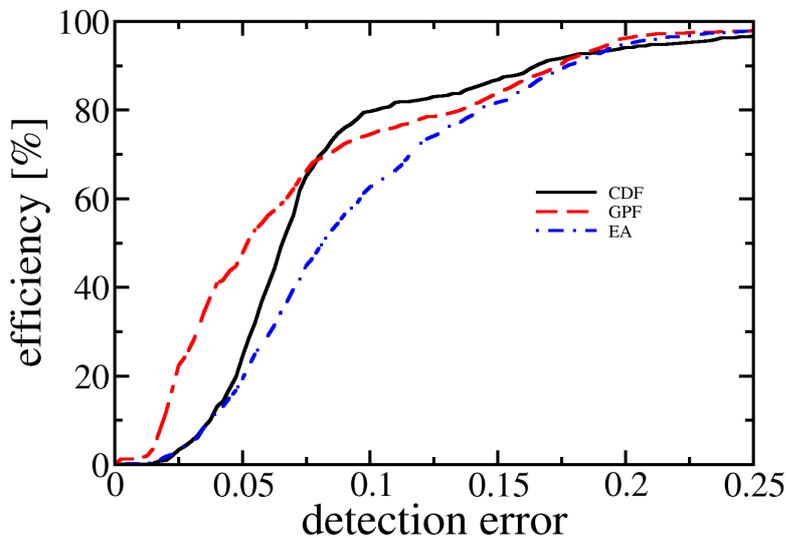

*Fig 11. The comparison of three algorithms for eye pupil location described in Sec. 3.*

In the range of low detection error ($d_{max} < 0.07$) the best results were obtained with GPF method (Sec 3.2). Then, the rapid grow of CDF algorithm efficiency is observed and up to the value of ($d_{max} < 0.15$) it remains the best one. EA method is the worst one in this range, but it catches up with the leading algorithms at $d_{max} > 0.15$. Over d_max=0.25 all the methods boast 100% efficiency.

| $d_{max}$ | CDF | GPF | EA |
|---|---|---|---|
| 0.02 | 1.0 % | 11.7 % | 1.8 % |
| 0.05 | 24.4 % | 47.7 % | 19.3 % |
| 0.1 | 79.7 % | 74.5 % | 62.6 % |
| 0.15 | 86.8 % | 83.8 % | 81.7 % |
| 0.2 | 94.0 % | 96.2 % | 94.9 % |
| 0.25 | 96.6 % | 97.9 % | 97.8 % |

Tab. 1. Eye pupil location algorithms efficiency at selected levels of the detection error $d_{max}$.

### 4.2. Comparison using webcam images

Although the performed tests are repeatable and provide objective quantitative results, the subjective appraisal of the algorithms by a user operating computer using eye-controlled interaction system could be completely different. Therefore we created the EyeTracker application, which is a part of eye driven interface and allows testing eye pupil location algorithms on static images taken from the BioID database. It requires Windows XP/Vista/7 operating system equipped with 32-bit version of MS Visual C++ runtime libraries [16]. The program can be downloaded from [17]. The main objective of this software is to enable computer operation based just on eye movement. The movement of eyes changes the position of mouse cursor, while blinking triggers clicking. The application settings allow for the selection of one of the described eye pupil location methods. Therefore users are given possibility to test and assess usefulness of the chosen algorithm in their own conditions.

In our work the comparison was performed using two VGA webcams, Philips SPC 900NC and Vimicro USB2.0 UVC. The SPC 900NC characterizes with better sharpness and overall image quality. All the three algorithms show similar good accuracy. The Vimicro webcam acquired much worse images. In this case the most efficient method seems to be CDF as it is not as much dependent on the image contrast as GPF or EA.

Another important factor for any real-time application is its performance. It was measured here in a number of processed frames per second. A laptop equipped with Intel C2D processor in a 320x240 mode enables all the algorithms to reach 15 fps, which is a limiting value for a webcam hardware and operating system drivers.

As a human-computer interface device the EyeTracker application is usable but it is hardly ergonomic. The main advantage of such a solution is lack of any requirements. All modern notebooks are equipped with sufficiently good webcams and necessary computing power.

## 5. Summary

We compared three algorithms for eye pupil location. Currently, all of them can be effectively used for gaze tracking and contactless computer operation. Although the other still lacks ergonomics, the technological progress will probably overcome that issue quickly. With better webcam images quality in terms of noise, sharpness and resolution, as well as growing computing power, operating computer using just a gaze will become as natural as using mouse, touchpad or touchscreen.